\RequirePackage{fix-cm}
\documentclass[smallextended]{svjour3}       
\smartqed  
\usepackage{amsmath}
\usepackage{graphicx,epsfig,wrapfig}
\usepackage{makeidx}
\usepackage{amsfonts}
\usepackage{amssymb}
\usepackage{amsmath}
\usepackage{dsfont}
\usepackage{mathrsfs}
\usepackage{slashed}
\usepackage{epstopdf}
\usepackage{bm}
\usepackage{color}
\usepackage[normalem]{ulem}
\usepackage{simplewick}

\newcommand{\ud}{\mathrm{d}}

\newcommand{\uvec}[1]{\boldsymbol{#1}}

\newcommand{\uind}[2]{^{#1}_{\phantom{#1}#2}}

\newcommand{\LRD}{\overset{\leftrightarrow}{D}\!\!\!\!\!\phantom{D}}

\begin{document}

\title{On the hadron mass decomposition}

\author{C\'edric Lorc\'e}

\institute{Centre de Physique Th\'eorique, \'Ecole polytechnique, 
	CNRS, Universit\'e Paris-Saclay, F-91128 Palaiseau, France\\\email{cedric.lorce@polytechnique.edu}}

\date{\today}

\maketitle

\begin{abstract}
We argue that the standard decompositions of the hadron mass overlook pressure effects, and hence should be interpreted with great care. Based on the semiclassical picture, we propose a new decomposition that properly accounts for these pressure effects. Because of Lorentz covariance, we stress that the hadron mass decomposition automatically comes along with a stability constraint, which we discuss for the first time. We show also that if a hadron is seen as made of quarks and gluons, one cannot decompose its mass into more than two contributions without running into trouble with the consistency of the physical interpretation. In particular, the so-called quark mass and trace anomaly contributions appear to be purely conventional. Based on the current phenomenological values, we find that in average quarks exert a repulsive force inside nucleons, balanced exactly by the gluon attractive force.
\end{abstract}

\section{Introduction}

According to the Standard Model of Particle Physics, the masses of almost all known elementary particles are generated through the Brout-Englert-Higgs (BEH) mechanism. The current light quark masses obtained from such a mechanism correspond however to only about 1\% of the nucleon mass. Therefore, the mass of the ordinary matter around us essentially finds its origin in the strong interactions which confine quarks and gluons inside hadrons, and not the BEH mechanism~\cite{Schumacher:2014jga,Gao:2015aax}.

Understanding the origin of the hadron mass in Quantum Chromodynamics (QCD) represents a formidable challenge owing to the relativistic, quantum and non-perturbative nature of the problem. The QCD lagrangian in the chiral limit appears to be scale invariant at the classical level. This implies in particular that all the hadron masses should vanish in that limit. Scale invariance is however an anomalous symmetry, in the sense that it appears to be broken at the quantum level by radiative corrections. This gives rise, via dimensional transmutation, to a dimensionful parameter $\Lambda_\text{QCD}\approx 0.2$ GeV in the theory~\cite{Callan:1970ze,Coleman:1973jx}, and therefore to a non-trivial spectrum. 

Lattice QCD calculation of hadron masses obtained from the analysis of correlations functions in Euclidean time is in remarkable agreement with the experimental spectrum~\cite{Durr:2008zz,Edwards:2012fx}. Unfortunately, this method gives little insight on how these masses arise from quark and gluon contributions. 

In order to address the question of the origin of the hadron mass, one should rather start from the energy-momentum tensor (EMT) of QCD. At the classical level, it reads
\begin{equation}
T^{\mu\nu}=\overline\psi\gamma^\mu \tfrac{i}{2}\LRD^\nu\psi-G^{a\mu\lambda}G^{a\nu}_{\phantom{a\nu}\lambda}+\tfrac{1}{4}\,\eta^{\mu\nu}G^2,
\end{equation}
where $\psi$ is the quark field, $G^{a\mu\nu}$ is the gluon field strength, $\LRD^\mu=\overset{\rightarrow}{\partial}\!\!\!\!\phantom{\partial}^\mu-\overset{\leftarrow}{\partial}\!\!\!\!\phantom{\partial}^\mu-2igA^{a\mu}t^a$ is the symmetric non-abelian covariant derivative, and $\eta_{\mu\nu}$ is the Minkowski metric. Note that, as argued e.g. in~\cite{Leader:2013jra,Lorce:2015lna}, there is no need in Particle Physics to require the EMT to be symmetric in its Lorentz indices. This aspect will however not affect the purpose of the present paper and will therefore not be discussed further. An important property of the total EMT is that it is conserved $\partial_\mu T^{\mu\nu}=0$, which is a consequence of the invariance of the theory under space-time translations. At the quantum level, it does not therefore require any global renormalization~\cite{Brown:1979pq}. 

The trace of the renormalized EMT is given by~\cite{Crewther:1972kn,Chanowitz:1972vd,Nielsen:1975ph,Adler:1976zt,Collins:1976yq,Nielsen:1977sy}
\begin{equation}\label{trace}
T^\mu_{\phantom{\mu}\mu}=\tfrac{\beta(g)}{2g}\,G^2+(1+\gamma_m)\,\overline\psi m\psi,
\end{equation}
where $m$ is the quark mass matrix and $\gamma_m$ is its anomalous dimension. The first term is known as the trace anomaly, which is a pure quantum effect as indicated by the $\beta$ function factor. In the chiral limit $m\to 0$, this term prevents the trace of the EMT to vanish, and hence prevents QCD to become a scale-invariant theory with trivial spectrum.

Mass decompositions can be obtained from the expectation value of the EMT. In the next Section, we briefly present the two standard decompositions used in the literature, and point out a couple of issues regarding their physical interpretation. To the best of our knowledge, some of these issues have not been addressed before, in particular regarding pressure effects. Based on the semiclassical picture, we propose in Section~\ref{sec3} a new decomposition free of these issues. Then we discuss in Section~\ref{sec4} the new picture in more detail and we estimate the various contributions using our current phenomenological knowledge. Finally, we summarize our results in Section~\ref{sec5}.

\section{A critique of the standard decompositions}\label{sec2}

\subsection{Trace decomposition}

From Poincar\'e invariance, the forward matrix elements of the total EMT in a hadron state with momentum $P$ and spin $j\leq\tfrac{1}{2}$ reads~\cite{Jaffe:1989jz}
\begin{equation}\label{param0}
\langle P|T^{\mu\nu}(0)|P\rangle=2P^\mu P^\nu,
\end{equation}
where the state is normalized according to 
\begin{equation}\label{statenorm}
\langle P'|P\rangle=2P^0\,(2\pi)^3\,\delta^{(3)}(\uvec P'-\uvec P).
\end{equation}
The hadron mass $M$ can then be defined via the trace of the EMT~\cite{Shifman:1978zn,Donoghue:1987av,Jaffe:1989jz,Donoghue:1992dd,Luke:1992tm,Kharzeev:1995ij,Shifman:1999mk,Bressani:2005bv}
\begin{equation}\label{massdef}
2M^2=\langle P|T^\mu_{\phantom{\mu}\mu}(0)|P\rangle.
\end{equation}
This expression is quite appealing owing to its manifest Poincar\'e invariance. In particular, it relies on the identity $P^2=M^2$, which defines mass as the square root of one of the Casimir operators associated with the Poincar\'e group.

Using the explicit expression~\eqref{trace} for the trace of the EMT operator
\begin{equation}\label{tracedec}
2M^2=\langle P|\tfrac{\beta(g)}{2g}\,G^2|P\rangle+\langle P|(1+\gamma_m)\,\overline\psi m\psi|P\rangle,
\end{equation}
it is tempting to conclude that most of the light hadron masses comes from gluons, based on the fact that the second term on the right hand side is known to be a rather small contribution (except for the pion)~\cite{Shifman:1978zn,Kharzeev:2002fm,Roberts:2016mhh,Roberts:2016vyn,Krein:2017usp}. Note that this partonic picture applies whenever the renormalization scale is much larger than $2$ GeV. For a renormalization scale below $0.5$ GeV, the picture may look quite different and closer to a non-relativistic picture where the (effective) constituents are quite massive and the (residual) binding energy is small and negative~\cite{Roberts:2016mhh,Roberts:2016vyn}.
\newline

The physical interpretation of the origin of hadron mass based on Eq.~\eqref{massdef} suffers however from a couple of caveats. Indeed, it seems a bit odd from a physical point of view that the definition of mass involves the trace of the EMT only at a single point. Moreover, the precise form of Eq.~\eqref{param0}, and hence of Eq.~\eqref{massdef}, depends on how hadron states are normalized. A proper physical interpretation should in principle not depend on our conventions for the normalization of states. It is therefore more natural to define mass via the normalized expectation value of some spatially extended operator
\begin{equation}
\langle O\rangle=\frac{\langle P|\int\ud^3r\,O(r)|P\rangle}{\langle P|P\rangle}.
\end{equation}
It is understood that such an expectation value contains implicitly the subtraction of the corresponding vacuum matrix element. Note in particular that the precise form of Eq.~\eqref{param0} was fixed by the requirement that $\langle T^{0\mu}\rangle=P^\mu$. It follows that the proper expectation value of the trace of the EMT reads
\begin{equation}\label{normmass}
\langle T^\mu_{\phantom{\mu}\mu}\rangle=M^2/P^0
\end{equation}
independently of how states are normalized. The presence of the $P^0$ factor indicates that this expectation value is frame dependent. 

In our opinion, frame dependence is not really a problem since physical quantities often find a simple interpretation only in some particular set of reference frames. For instance, using the light-front form of dynamics~\cite{Brodsky:1997de}, Eq.~\eqref{normmass} becomes $\langle T^\mu_{\phantom{\mu}\mu}\rangle=M^2/P^+$ which can be interpreted in the symmetric frame (defined by $\uvec P_\perp=\uvec 0_\perp$) as twice the light-front energy $\langle T^\mu_{\phantom{\mu}\mu}\rangle|_{\uvec P_\perp=\uvec 0_\perp}=2P^-$. Frame dependence\footnote{Note that in the light-front form of dynamics, boost invariance is sometimes incorrectly called frame independence.} seems to be a generic feature associated with many decompositions of a system. If one insists on manifest Lorentz invariance\footnote{Sometimes people consider that physical quantities have to be Lorentz invariant, and so they conclude that decompositions are often unphysical and hence uninteresting or irrelevant. Interestingly, this seems to be the dominant thought regarding diffeomorphism invariance in General Relativity, but alternative descriptions based on the concept of vielbein are more flexible on the matter. We feel that this is too strict a requirement, for that many measurable quantities like e.g. energy and spin would be considered as unphysical.}, it turns out that Lorentz non-invariant quantities defined in a given frame can be formally expressed as Lorentz scalars~\cite{Leader:2013jra,Hoodbhoy:1998bt}. By construction, the latter will give the same result in any frame, and will coincide with the original non-invariant quantities in the given frame. Typical examples are the proper time and length which are Lorentz-invariant quantities giving respectively the time and length measured in the rest frame of the system. In other words, it is in principle always possible to express quantities in a Lorentz-invariant form, but the price to pay is that the physical interpretation necessarily singles out a privileged frame. In the present case, the frame dependence of Eq.~\eqref{normmass} originates from the fact that we are considering the operator $\int\ud^3r\,T^\mu_{\phantom{\mu}\mu}(r)$, which is not Lorentz invariant unlike $T^\mu_{\phantom{\mu}\mu}(0)$. It is however in line with the standard expectation that mass effects in hadronic matrix elements should become negligible in the ultra-relativistic regime $P^0\gg M$. If one insists on preserving manifest Lorentz invariance, one can alternatively consider the integral over the proper volume $\frac{P^0}{M}\int\ud^3r\,T^\mu_{\phantom{\mu}\mu}(r)$. Frame dependence will disappear from the formal expressions but it will remain in their physical interpretation.

Beside the question of frame dependence, a more important point we would like to stress is that the physical interpretation should be directly based on the quantum operator. Providing an interpretation at the level of the matrix element may be misleading. For example, in Eq.~\eqref{massdef} the hadron mass is connected to the trace of the EMT at the level of matrix elements, but the connection remains obscure at the operator level, especially when the operator is decomposed into several contributions. Indeed, the trace $T\uind{\mu}{\mu}$ involves beside energy $T\uind{0}{0}$ the spatial components $T\uind{i}{i}$ usually associated with normal stresses. It seems also counterintuitive that the contribution from the quark mass to the hadron mass is not proportional to $m$, but rather to $\sqrt{m}$. In fact, there exist (infinitely) \emph{many} operators whose forward matrix elements are proportional to some power of the hadron mass. Following the same logic as for the trace operator, one could then in principle argue that these operators can be used to provide alternative definitions (and hence alternative decompositions) of the hadron mass, raising the question of deciding which one is the ``correct'' one. This happens simply because the hadron mass gives the natural scale of the matrix elements. The only way out is to use a quantum operator with a clear connection to the concept of mass.

In view of all these caveats, we feel that claims about the physical origin of hadron mass based on Eq.~\eqref{massdef} are somewhat misleading. Note also that the above remarks can similarly be applied to the Gell-Mann--Oakes--Renner formula~\cite{GellMann:1968rz}.

\subsection{Ji's decomposition}

About 20 years ago, Ji proposed a decomposition of the hadron mass analogous to the virial theorem for a harmonic oscillator and the hydrogen atom~\cite{Ji:1994av,Ji:1995sv}, see also~\cite{Gaite:2013uqa}. As a first step, he decomposed the QCD EMT into traceless and trace parts
\begin{equation}\label{barhat}
T^{\mu\nu}=\bar T^{\mu\nu}+\hat T^{\mu\nu}
\end{equation}
with $\hat T^{\mu\nu}=\tfrac{1}{4}\,\eta^{\mu\nu}T^\alpha_{\phantom{\alpha}\alpha}$. This choice is of course not unique\footnote{The general trace part is defined as $\hat T^{\mu\nu}_c=\hat T^{\mu\nu}+c\,\bar T^{\mu\nu}$ with $c$ an arbitrary constant.}, but it is well suited for treating the trace anomaly contribution. Using the relativistic normalization~\eqref{statenorm} for the states, the corresponding forward matrix elements read
\begin{align}
\langle P|\bar T^{\mu\nu}(0)|P\rangle&=2\left(P^\mu P^\nu-\tfrac{1}{4}\,\eta^{\mu\nu}M^2\right),\\
\langle P|\hat T^{\mu\nu}(0)|P\rangle&=\tfrac{1}{2}\,\eta^{\mu\nu}M^2,
\end{align}
and are renormalization scale independent, as required by Lorentz symmetry. In the Hamiltonian formalism, the mass of a system is defined as the total energy in the rest frame. It then follows that
\begin{equation}\label{virial}
\langle\bar T^{00}\rangle|_{\uvec P=\uvec 0}=\tfrac{3}{4}\,M,\qquad
\langle\hat T^{00}\rangle|_{\uvec P=\uvec 0}=\tfrac{1}{4}\,M.
\end{equation}
In other words, the trace of the EMT appears to contribute only to 25\% of the hadron mass.

Ji proceeded with a decomposition of the traceless part into quark and gluon contributions, and of the trace part into quark mass and trace anomaly contributions
\begin{equation}\label{Tmunudec}
T^{\mu\nu}=\bar T^{\mu\nu}_q+\bar T^{\mu\nu}_g+\hat T^{\mu\nu}_m+\hat T^{\mu\nu}_a.
\end{equation}
The corresponding individual forward matrix elements are not protected by Lorentz symmetry
\begin{align}
\langle P|\bar T^{\mu\nu}_q(0)|P\rangle&=2\,a(\mu^2)\left(P^\mu P^\nu-\tfrac{1}{4}\,\eta^{\mu\nu}M^2\right),\label{Tq}\\
\langle P|\bar T^{\mu\nu}_g(0)|P\rangle&=2\,[1-a(\mu^2)]\left(P^\mu P^\nu-\tfrac{1}{4}\,\eta^{\mu\nu}M^2\right),\\
\langle P|\hat T^{\mu\nu}_m(0)|P\rangle&=\tfrac{1}{2}\,b(\mu^2)\,\eta^{\mu\nu}M^2,\\
\langle P|\hat T^{\mu\nu}_a(0)|P\rangle&=\tfrac{1}{2}\,[1-b(\mu^2)]\,\eta^{\mu\nu}M^2,\label{Ta}
\end{align}
and so the coefficients $a(\mu^2)$ and $b(\mu^2)$ depend generally on the renormalization scale $\mu$. It follows from Eq.~\eqref{Tmunudec} that the Hamiltonian operator can naturally be decomposed as
\begin{equation}
H=H'_q+H_g+H'_m+H_a,
\end{equation}
where
\begin{align}
H'_q&=\int\ud^3r\,\bar T^{00}_q(r),\label{Hq}\\
H_g&=\int\ud^3r\,\bar T^{00}_g(r),\\
H'_m&=\int\ud^3r\,\hat T^{00}_m(r),\\
H_a&=\int\ud^3r\,\hat T^{00}_a(r).\label{Ha}
\end{align}
Owing to the QCD equations of motion, the quark contribution can be put in the form\footnote{Strictly speaking, the equality holds only at the level of matrix elements}
\begin{equation}
H'_q=\int\ud^3r\left[\psi^\dag(-i\uvec D\cdot\uvec\alpha)\psi+\tfrac{3}{4}\,\overline\psi m\psi\right],
\end{equation}
where $\uvec D=\uvec\nabla+ig\uvec A$. Rearranging a bit the quark mass term between $H'_q$ and $H'_m$, 
\begin{equation}
H'_q+H'_m=H_q+H_m
\end{equation}
with
\begin{align*}
H_q&=\int\ud^3r\,\psi^\dag(-i\uvec D\cdot\uvec\alpha)\psi,\\
H_m&=\int\ud^3r\left(1+\tfrac{1}{4}\,\gamma_m\right)\overline\psi m\psi,
\end{align*}
Ji arrived at the following decomposition of the hadron mass in the rest frame
\begin{equation}\label{totalmass}
M=M_q+M_g+M_m+M_a,
\end{equation}
where
\begin{equation}
M_i=\left.\frac{\langle P|H_i|P\rangle}{\langle P|P\rangle}\right|_{\uvec P=\uvec 0}\qquad i=q,g,m,a.
\end{equation}
In particular, using the parametrization in Eqs.~\eqref{Tq}-\eqref{Ta}, the individual contributions can be expressed as
\begin{align}
M_q&=\tfrac{3}{4}\left(a-\tfrac{b}{1+\gamma_m}\right)M,\label{Mq}\\
M_g&=\tfrac{3}{4}\left(1-a\right)M,\\
M_m&=\tfrac{1}{4}\,\tfrac{4+\gamma_m}{1+\gamma_m}\,b\,M,\\
M_a&=\tfrac{1}{4}\left(1-b\right)M.\label{Ma}
\end{align}
In the chiral limit, one has $b=0$ and the gluon energy arising from the trace anomaly corresponds precisely to the vacuum energy introduced phenomenologically in the MIT bag model~\cite{Chodos:1974je}. This decomposition is quite popular in hadronic physics~\cite{Gao:2015aax} and motivated many lattice QCD calculations, see e.g.~\cite{Meyer:2007tm,Yang:2014xsa,Bali:2016lvx} for recent related works.
\newline

Ji's decomposition is sometimes criticized because it is performed in a specific frame and applies to massive states only~\cite{Roberts:2016mhh,Roberts:2016vyn}. As we already argued in the previous section, the frame dependence is not really a problem but a general feature. Note also that Ji's decomposition can formally be put in a covariant form by considering the Lorentz-invariant quantity $\langle T^{0\mu}u_\mu\rangle$, where $u^\mu\equiv P^\mu/M$ is interpreted as the hadron four-velocity. Although this form is frame independent, its physical interpretation becomes simple only in the rest frame, where $u^\mu=(1,\uvec 0)$. For a massless state, since there is no rest frame, one can consider instead the energy decomposition in any frame. In this case, there is no contribution from the trace part since $P^2=0$ and so only the coefficient $a(\mu^2)$ is needed\footnote{Note that one could still formally use Ji's decomposition based on the covariant quantities $\lim_{P^2\to 0}\langle T^{0\mu}P_\mu\rangle/P^2$ or $\partial\langle T^{0\mu}P_\mu\rangle/\partial P^2|_{P^2=0}$, which are in the spirit of Ji's remark~\cite{Ji:1995sv} that although the {\it overall scale} is essentially determined by the anomaly in the light hadron sector, the {\it relative magnitudes} of the various contributions reflect essential aspects of the underlying quark-gluon dynamics in the non-perturbative regime.}.

The actual problem with Ji's decomposition is that the separation of the EMT into traceless and trace parts is not inconsequential for the physical interpretation of the individual contributions\footnote{Paying attention not to introduce spurious contributions that sum up to zero in a decomposition is a general problem. For example, the proper definition of angular momentum at the level of spatial distribution has recently been discussed in detail in~\cite{Lorce:2017wkb}, where the problem was solved by treating with care all contributions that vanish under integration.}. Namely, although $T^{00}$, $\bar T^{00}$ and $\hat T^{00}$ all have the dimension of energy densities, they actually correspond to different thermodynamic potentials as we will show in the following. By focusing on the $\mu=\nu=0$ component in the rest frame, Ji's decomposition does not make any distinction between the Lorentz tensors $P^\mu P^\nu$ and $M^2\eta^{\mu\nu}$, and therefore disregards pressure effects.

To sum up, although all terms in Ji's decomposition~\eqref{totalmass} can formally be defined and evaluated, they cannot however be interpreted as pure mass contributions. In the next section, we will explain how to achieve a proper mass decomposition based on a more covariant treatment using the semiclassical picture.

\section{A new decomposition}\label{sec3}

Consider some decomposition of the EMT at the operator level $T^{\mu\nu}=\sum_iT^{\mu\nu}_i$. From the general parametrization of the matrix element of a rank-two tensor in a state with mass $M$ and spin $j\leq\tfrac{1}{2}$~\cite{Ji:1996ek,Bakker:2004ib,Leader:2013jra}, we obtain in the forward limit
\begin{equation}\label{param}
\langle P| T^{\mu\nu}_i(0)|P\rangle=2P^\mu P^\nu A_i(0)+2M^2\eta^{\mu\nu}\bar C_i(0),
\end{equation}
where $A_i(0)$ and $\bar C_i(0)$ are two energy-momentum form factors evaluated at zero momentum transfer and depending on the renormalization scale $\mu$. They are related to Ji's coefficients as follows
\begin{equation}\label{dictionary}
a_i=A_i(0),\qquad b_i=A_i(0)+4\,\bar C_i(0).
\end{equation}
Four-momentum conservation implies that
\begin{equation}\label{sumrule}
\sum_iA_i(0)=1,\qquad \sum_i\bar C_i(0)=0,
\end{equation} 
as one can see from Eq.~\eqref{param0}.

In order to identify the physical meaning of $T^{\mu\nu}_i$, we consider its expectation value averaged over the hadron volume $\mathcal V=M V/P^0$, with $V$ denoting the hadron proper volume,
\begin{equation}
\frac{1}{\mathcal V}\,\langle T^{\mu\nu}_i\rangle=\frac{\langle P|T^{\mu\nu}_i(0)|P\rangle}{2MV}.
\end{equation}
In the hadron rest frame, we find
\begin{equation}\label{Thetamunu}
\frac{1}{\mathcal V}\,\langle T^{\mu\nu}_i\rangle|_{\uvec P=\uvec 0}=\begin{pmatrix}\varepsilon_i&&&\\&p_i&&\\
&&p_i&\\
&&&p_i\end{pmatrix},
\end{equation}
where
\begin{equation}
\varepsilon_i\equiv[A_i(0)+\bar C_i(0)]\,\tfrac{M}{V}\qquad\text{and}\qquad p_i\equiv-\bar C_i(0)\,\tfrac{M}{V}
\end{equation}
are naturally interpreted as partial proper (internal) energy density and isotropic pressure, respectively. Note that $\varepsilon_i$ and $p_i$ are average quantities over the hadron proper volume, and hence are proportional to the average proper energy density of the hadron $M/V$. In a generic frame, we can write
\begin{equation}
\frac{1}{\mathcal V}\,\langle T^{\mu\nu}_i\rangle=(\varepsilon_i+p_i)u^\mu u^\nu-p_i\,\eta^{\mu\nu}
\end{equation}
which has the same structure as the EMT of an element of perfect fluid in relativistic hydrodynamics~\cite{Eckart:1940te}. Using now the relation~\eqref{dictionary}, we find that Ji's coefficients correspond to the notions of enthalpy density $h_i$ and interaction measure $I_i$
\begin{equation}\label{aint}
h_i=\varepsilon_i+p_i=a_i\,\tfrac{M}{V},\qquad
I_i=\varepsilon_i-3p_i=b_i\,\tfrac{M}{V}.
\end{equation}
Remembering that $\varepsilon_i$ and $p_i$ are average quantities over the proper volume, the thermodynamic potentials are then simply defined as
\begin{equation}
U_i=\varepsilon_iV=[A_i(0)+\bar C_i(0)]\,M,\qquad H_i=h_iV=A_i(0)\,M,
\end{equation}
where $U_i$ is the partial internal energy, $H_i=U_i+W_i$ is the partial enthalpy, and $W_i=p_iV$ is the partial pressure-volume work. The latter quantity represents the energy associated with the occupation of the volume $V$ by the subsystem $i$ pushing ($p_i>0$) or pulling ($p_i<0$) the rest of the system. 

Since the internal energy represents the sum of all the kinetic and potential energies, the proper mass decomposition we were looking for is
\begin{equation}
M=\sum_i U_i.
\end{equation}
Moreover, the total pressure-volume work should vanish for a stable system 
\begin{equation}
\sum_i W_i=0.
\end{equation}
These two conditions are naturally equivalent to Eq.~\eqref{sumrule}.

\section{Discussion}\label{sec4}

In the previous section, we observed that the generic form of the EMT for a hadron with spin $j\leq\tfrac{1}{2}$ is characterized by two Lorentz scalar quantities similarly to that of an element of perfect fluid. We are of course not claiming that hadrons consist of a set of perfect fluids\footnote{Similarly, scalar fields in General Relativity are often mimicked by effective perfect fluids, see e.g.~\cite{Madsen:1988ph,Faraoni:2012hn}.}, let alone that a hydrodynamical description is quite difficult to justify in this context. Forward matrix elements of the EMT, like the ones we considered, allow us to determine only \emph{static} mechanical properties of the state. In the semiclassical limit, these matrix elements can effectively be thought of as a continuum description averaged over time~\cite{Chodos:1974je,Polyakov:2002wz,Polyakov:2002yz}. We simply adopted the terminology of Continuum Mechanics to identify the physical meaning of the various energy-momentum form factors. 

Decomposing the EMT into various contributions $T^{\mu\nu}=\sum_iT^{\mu\nu}_i$ amounts in practice to an \emph{effective} (time-averaged) coupled multifluid picture of the hadron~\cite{Rezzolla:2013}, where each species of the mixture is regarded as a separate continuum coexisting with the continuums made up of other species. Each continuum is then described by its own (partial) energy density $\varepsilon_i$ and pressure $p_i$. Since these quantities are averaged over the hadron proper volume, they trivially obey a barotropic equation of state for species $i$
\begin{equation}
p_i=w_i\,\varepsilon_i.
\end{equation}
Depending on the value of the parameter $w_i$, this equation of state describes an element of effective perfect fluid from phantom to ekpyrotic matter~\cite{Saha:2009cj,Delsate:2012ky}
\begin{equation}
\begin{aligned}
w_i&<-1&\text{phantom matter}\\
w_i&=-1&\text{cosmological constant}\\
w_i&\in(-1,-\tfrac{1}{3})&\text{quintessence}\\
w_i&=0&\text{dust}\\
w_i&=\tfrac{1}{3}&\text{radiation}\\
w_i&\in(\tfrac{1}{3},1)&\text{hard Universe}\\
w_i&=1&\text{stiff matter}\\
w_i&>1&\text{ekpyrotic matter}
\end{aligned}
\end{equation}

\subsection{Back to the old decompositions}

Considering the trace of Eq.~\eqref{Thetamunu} shows that the trace decomposition in Eq.~\eqref{tracedec} does not represent a decomposition of the total energy of the system, but rather a decomposition of the interaction measure $M=\sum_i(U_i-3W_i)=\sum_i(\varepsilon_i-3p_i)V$. Each individual contribution involves, beside internal energy $U_i$, the partial pressure-volume work $W_i$. For a stable system, the total pressure-volume work $\sum_iW_i=0$ has to vanish, explaining why one obtains at the end just the total mass $M$ of the system\footnote{In principle, one can construct infinitely many decompositions of the form $M=\sum_i(U_i+c W_i)$ with $c$ an arbitrary constant.}. 

Except possibly for the pion, the first term on the right hand side of Eq.~\eqref{tracedec} corresponding to the gluon contribution is expected to dominate in light hadrons. Contrary to what is often claimed in the literature~\cite{Shifman:1978zn,Kharzeev:2002fm,Roberts:2016mhh,Roberts:2016vyn,Krein:2017usp}, this \emph{does not} mean however that gluons are at the origin of most of the light hadron masses. It actually indicates that
\begin{equation}
(\varepsilon_g-3p_g)\gg |\varepsilon_q-3p_q|.
\end{equation}
In non-relativistic systems, one can safely neglect pressures in front of energy densities, leading then to $\varepsilon_g\gg |\varepsilon_q|$, but in relativistic systems like light hadrons, one may expect that pressures become comparable to energy densities. Since stability of the system imposes $p_q=-p_g$, assuming an approximate equal sharing of the energy between quarks and gluons $\varepsilon_q\approx\varepsilon_g>0$ leads to the conclusion that
\begin{equation}
p_q>0,\qquad p_g<0.
\end{equation}
We will see later that such a scenario is indeed supported by the current phenomenology. So, in average, quarks are responsible for the repulsive force and gluons for the attractive force inside light hadrons. The two antagonist forces balance each other so as to lead to a stable bound state. This is reminiscent of the structure of stars, where the repulsive force caused by the gas pressure is balanced by the attractive gravitational force\footnote{Note that in the context of teleparallel gravity, the concepts of gravitational energy and pressure can be defined, see e.g.~\cite{Maluf:2005kn,Maluf:2012na,Castello-Branco:2013iza}.}. 
\newline

Hadrons are bound states of two types of constituents, namely quarks and gluons. We can therefore decompose the QCD EMT into quark and gluon contributions
\begin{equation}
T^{\mu\nu}=T^{\mu\nu}_q+T^{\mu\nu}_g,
\end{equation}
where
\begin{align}
T^{\mu\nu}_q&=\overline\psi\gamma^\mu \tfrac{i}{2}\LRD^\nu\psi+\tfrac{1}{4}\,\eta^{\mu\nu}\gamma_m\,\overline\psi m\psi,\label{Tqmunu}\\
T^{\mu\nu}_g&=-G^{a\mu\lambda}G^{a\nu}_{\phantom{a\nu}\lambda}+\tfrac{1}{4}\,\eta^{\mu\nu}\left(1+\tfrac{\beta(g)}{2g}\right)G^2\label{Tgmunu},
\end{align}
although there is some arbitrariness in doing so owing to quark-gluon interactions~\cite{Leader:2013jra}. Here, the quark mass term is part of $T^{\mu\nu}_q$ and the trace anomaly is part of $T^{\mu\nu}_g$. Accordingly, it seems therefore natural to adopt an effective coupled two-fluid picture where $T^{\mu\nu}_{q,g}$ are treated as separate EMT and hence described in terms of their own energy densities $\varepsilon_{q,g}$ and pressures $p_{q,g}$. 

Let us now come back to Ji's decomposition. Looking at the expressions~\eqref{Mq}-\eqref{Ma}, we observe that the individual contributions to the hadron mass correspond to different linear combinations of the coefficients $a$ and $b$. In the effective coupled two-fluid picture, they correspond therefore to different linear combinations of energy density and pressure owing to Eq.~\eqref{aint}. This is of course not physically acceptable, in the sense that one is adding apples and oranges. It is however mathematically correct since the sum over species gives at the end just the hadron mass $M$, once again thanks to the stability constraint $\sum_iW_i=0$. The only way to make physical sense out of Ji's decomposition within the effective coupled two-fluid picture is to define the quark mass and trace anomaly contributions $\varepsilon_{m,a}$ from the onset, and to split the quark and gluon energies as follows
\begin{equation}
\varepsilon_q=\varepsilon_{q-m}+\varepsilon_m,\qquad \varepsilon_g=\varepsilon_{g-a}+\varepsilon_a,
\end{equation}
with $\varepsilon_{q-m}\equiv\varepsilon_q-\varepsilon_m$ and $\varepsilon_{g-a}\equiv\varepsilon_g-\varepsilon_a$, while keeping the pressure part unchanged. This is consistent, but the price to pay is a loss of covariance since different components of the EMT are not treated in the same way.

What Ji did actually in Eq.~\eqref{Tmunudec} is a covariant decomposition of the QCD EMT into \emph{four} parts. He considered with Eqs.~\eqref{Hq}-\eqref{Ha} that each quantity $\bar T^{00}_q$, $\bar T^{00}_g$, $\hat T^{00}_m$, and $\hat T^{00}_a$ should represent an energy density, and hence \emph{implicitly} adopted an effective coupled four-fluid picture of the hadron. The problem with such a description is that e.g. gluons carying kinetic and potential energies are considered as \emph{different} from those involved in the trace anomaly, in the sense that they are effectively treated as two distinct continuums with their own equations of state. Our opinion is that this is not acceptable from a physical point of view. Since hadrons are composed of quarks and gluons, an effective description of hadrons in terms of only two continuums is more natural.

\subsection{Virial decomposition}

In the effective coupled two-fluid picture, the hadron mass is decomposed into quark and gluon internal energies $M=U_q+U_g$, where $U_{q,g}=\langle T^{00}_{q,g}\rangle|_{\uvec P=\uvec 0}$ with $T^{\mu\nu}_{q,g}$ given by Eqs.~\eqref{Tqmunu} and \eqref{Tgmunu}. The question now is whether one can further decompose into parton kinetic and potential energies, quark mass and trace anomaly contributions, similarly to Ji's decomposition.

Let us start with the gluon sector. Since the only contribution to the trace of the gluon EMT comes from the trace anomaly, it seems natural to write
\begin{equation}
T^{\mu\nu}_g=\bar T^{\mu\nu}_g+\hat T^{\mu\nu}_g.
\end{equation}
For the quark sector, we reshuffle the quark mass terms after the separation into traceless and trace parts
\begin{equation}
T^{\mu\nu}_q=\tilde T^{\mu\nu}_q+\check T^{\mu\nu}_q,
\end{equation}
where
\begin{equation}\label{reshuffling}
\tilde T^{\mu\nu}_q=\bar T^{\mu\nu}_q-c \,\hat T^{\mu\nu}_q,\qquad
\check T^{\mu\nu}_q=\left(1+c\right)\hat T^{\mu\nu}_q.
\end{equation}
The constant $c$ is then chosen such that $\tilde T^{00}_q$ takes the form $\psi^\dag(-i\uvec D\cdot\uvec\alpha)\psi$ upon using the QCD equations of motion, leading us to
\begin{equation}
c=\tfrac{3}{1+\gamma_m}.
\end{equation}

Now, contrary to Ji, we will not treat $\bar T^{\mu\nu}_i$ and $\hat T^{\mu\nu}_i$ as separate EMT, but rather as mere parts of the EMT $T^{\mu\nu}_i$. We then have 
\begin{equation}
\langle\bar T^{00}_i\rangle|_{\uvec P=\uvec 0}=\tfrac{3}{4}\left(U_i+W_i\right),\qquad
\langle\hat T^{00}_i\rangle|_{\uvec P=\uvec 0}=\tfrac{1}{4}\left(U_i-3W_i\right).
\end{equation}
In other words, we see that traceless parts contribute to $\tfrac{3}{4}$ and trace parts to $\tfrac{1}{4}$ of the internal energy content. Summing over all the species $i$ naturally leads to Eq.~\eqref{virial}. Keeping only the internal energy contributions leads to the following finer decomposition of the hadron mass in the effective coupled two-fluid picture
\begin{equation}\label{finerU}
M=\tilde U_q+\check U_q+\bar U_g+\hat U_g
\end{equation}
with
\begin{align}
\tilde U_q=\tfrac{3}{4}\,\tfrac{\gamma_m}{1+\gamma_m}\,U_q,\qquad \check U_q=\tfrac{1}{4}\,\tfrac{4+\gamma_m}{1+\gamma_m}\,U_q,\qquad \bar U_g=\tfrac{3}{4}\,U_g,\qquad \hat U_g=\tfrac{1}{4}\,U_g.
\end{align}
For comparison, Ji's decomposition reads explicitly
\begin{align}
M_q&=\langle \tilde T^{00}_q\rangle|_{\uvec P=\uvec 0}=\tilde U_q+\tfrac{3}{4}\,\tfrac{4+\gamma_m}{1+\gamma_m}\,W_q,\\
M_m&=\langle \check T^{00}_q\rangle|_{\uvec P=\uvec 0}=\check U_q-\tfrac{3}{4}\,\tfrac{4+\gamma_m}{1+\gamma_m}\,W_q,\\
M_g&=\langle \bar T^{00}_g\rangle|_{\uvec P=\uvec 0}=\bar U_g+\tfrac{3}{4}\,W_g,\\
M_a&=\langle \hat T^{00}_g\rangle|_{\uvec P=\uvec 0}=\hat U_g-\tfrac{3}{4}\,W_g,
\end{align}
and differs from our finer decomposition by a reshuffling of the energy contributions within the quark and gluon sectors, separately.

We refrain from interpreting $\check U_q$ as quark mass contribution and $\hat U_g$ as trace anomaly contribution. Indeed, in the classical limit $\gamma_m\to 0$ we have $\check U_q=U_q$ which should also include quark kinetic and potential energies. Moreover, the contribution $\hat U_g$ does not vanish when the trace anomaly is set to zero $\langle\hat T^{\mu\nu}_g\rangle=0$. By keeping only the internal energy contributions, we lost the direct connection with the matrix elements and hence a simple physical interpretation. At best, our finer hadron mass decomposition~\eqref{finerU} can be seen as some sort of virial decomposition~\cite{Ji:1994av,Ji:1995sv,Gaite:2013uqa}. 
\newline

A similar treatment can be applied to the pressure-volume work. In the effective coupled two-fluid picture, the partial pressure-volume works of quarks and gluons satisfy $0=W_q+W_g$, where $W_{q,g}=\langle T^{33}_{q,g}\rangle|_{\uvec P=\uvec 0}$. Treating once again $\bar T^{\mu\nu}_i$ and $\hat T^{\mu\nu}_i$ as separate EMT, we get 
\begin{equation}
\langle\bar T^{33}_i\rangle|_{\uvec P=\uvec 0}=\tfrac{1}{4}\left(W_i+U_i\right),\qquad\langle\hat T^{33}_i\rangle|_{\uvec P=\uvec 0}=\tfrac{1}{4}\left(3W_i-U_i\right).
\end{equation}
In this case, traceless parts contribute to $\tfrac{1}{4}$ and trace parts to $\tfrac{3}{4}$ of the pressure-volume work. Summing over all the species $i$, we obtain the analogue of Eq.~\eqref{virial} for pressure-volume work
\begin{equation}
\langle\bar T^{33}\rangle|_{\uvec P=\uvec 0}=\tfrac{1}{4}\,M,\qquad\langle\hat T^{33}_i\rangle|_{\uvec P=\uvec 0}=-\tfrac{1}{4}\,M.
\end{equation}
Keeping only the pressure-volume work contributions leads to the following finer decomposition of the hadron stability constraint in the effective coupled two-fluid picture
\begin{equation}\label{finerW}
0=\tilde W_q+\check W_q+\bar W_g+\hat W_g
\end{equation}
with
\begin{align}
\tilde W_q=-\tfrac{1}{4}\,\tfrac{8-\gamma_m}{1+\gamma_m}\,W_q,\qquad \check W_q=\tfrac{3}{4}\,\tfrac{4+\gamma_m}{1+\gamma_m}\,W_q,\qquad \bar W_g=\tfrac{1}{4}\,W_g,\qquad \hat W_g=\tfrac{3}{4}\,W_g.
\end{align}
For comparison, a similar decomposition in the effective coupled four-fluid picture reads
\begin{equation}\label{Jipressure}
0=\mathcal W_q+\mathcal W_m+\mathcal W_g+\mathcal W_a
\end{equation}
with
\begin{align}
\mathcal W_q&=\langle \tilde T^{33}_q\rangle|_{\uvec P=\uvec 0}=\tilde W_q+\tfrac{1}{4}\,\tfrac{4+\gamma_m}{1+\gamma_m}\,U_q,\\
\mathcal W_m&=\langle \check T^{33}_q\rangle|_{\uvec P=\uvec 0}=\check W_q-\tfrac{1}{4}\,\tfrac{4+\gamma_m}{1+\gamma_m}\,U_q,\\
\mathcal W_g&=\langle \bar T^{33}_g\rangle|_{\uvec P=\uvec 0}=\bar W_g+\tfrac{1}{4}\,U_g,\\
\mathcal W_a&=\langle \hat T^{33}_g\rangle|_{\uvec P=\uvec 0}=\hat W_g-\tfrac{1}{4}\,U_g,
\end{align}
and differs from our finer decomposition by a reshuffling of the work contributions within the quark and gluon sectors, separately. Once again, we refrain from interpreting $\check W_q$ as quark mass contribution and $\hat W_g$ as trace anomaly contribution. At best, Eq.~\eqref{finerW} can be seen as some sort of virial decomposition.
\newline

The above finer decompositions are not based on the nature of the constituents but on Lorentz symmetry and this has a dramatic impact on the physical picture. Indeed, we were not able find a simple physical interpretation for the individual contributions to the decompositions~\eqref{finerU} and \eqref{finerW}, because the effective coupled two-fluid picture forced us to discard some of the terms in the matrix elements for consistency. In order to avoid this, one has to adopt a picture where the number of effective fluids is at least equal to the number of contributions. In the present case, we need at least an effective four-fluid picture, leading us back directly to Ji's decomposition. We have however already argued that the four-fluid picture is not natural from the physical point of view, since gluons carrying kinetic and potential energies are the same as those involved in the trace anomaly. Similarly, quarks carrying kinetic and potential energies are the same as those characterized by mass $m$.

Another indication that a decomposition based on Lorentz symmetry is not natural from the physical point of view is the following. Any EMT with the structure~\eqref{Thetamunu} and generic equation of state $p_i=w_i\,\varepsilon_i$ can be represented as a particular combination of two EMT with \emph{fixed} equations of state. This can be seen as some sort of decomposition onto a basis. For example, through a decomposition of the EMT into traceless and trace parts $T^{\mu\nu}_i=\bar T^{\mu\nu}_i+\hat T^{\mu\nu}_i$, the set $\{\varepsilon_i,p_i\}$ can formally be replaced by $\{\bar\varepsilon_i,\hat\varepsilon_i\}$ with $\bar\varepsilon_i=\tfrac{3}{4}\left(\varepsilon_i+p_i\right)$ and $\hat\varepsilon_i=\tfrac{1}{4}\left(\varepsilon_i-3p_i\right)$, since traceless EMT $\bar T^{\mu\nu}_i$ are characterized by the equation of state $\bar p_i=\tfrac{1}{3}\,\bar\varepsilon_i$ and pure trace EMT $\hat T^{\mu\nu}_i$ are characterized by $\hat p_i=-\hat\varepsilon_i$. In Ji's decomposition, the gluon contribution is indeed divided into kinetic and potential energies treated as a pure radiation $w_g=\tfrac{1}{3}$, and trace anomaly treated as a cosmological constant $w_a=-1$. Since the choice of a basis is not unique, the decomposition based on Lorentz symmetry is purely conventional, and hence artificial. In fact, Lorentz symmetry has \emph{already} been used to provide a physical interpretation of the various components of the EMT, namely by distinguishing in our case energy density $\varepsilon_i$ from pressure $p_i$ in Eq.~\eqref{Thetamunu}. One can then easily understand why using Lorentz symmetry \emph{again} to perform a decomposition amounts to choosing an arbitrary basis of EMT with fixed equations of state not determined by the physics of the problem.

To summarize, although a decomposition of the hadron mass or pressure-volume work into four contributions can formally be achieved in the effective coupled two-fluid picture, the individual terms cannot be interpreted in a simple way. If the hadron is seen as made of quarks and gluons, the only unambiguous decompositions are
\begin{equation}\label{newdecres}
M=U_q+U_g,\qquad 0=W_q+W_g.
\end{equation}
Any further decomposition based on Lorentz symmetry, like e.g. separation of traceless and trace parts, is artificial and purely arbitrary. The only physically acceptable further decompositions are those based on constituent properties, like e.g. flavor or polarization~\cite{Lorce:2014mxa,Bhoonah:2017olu}.

\subsection{Phenomenology}

All that is needed to characterize the quark and gluon EMT~\eqref{param} are the form factors $A_{q,g}(0)$ and $\bar C_{q,g}(0)$ in the forward limit. Using the energy-momentum sum rules~\eqref{sumrule}, we can reduce this set to e.g. the quark form factors $A_q(0)$ and $\bar C_q(0)$, or equivalently Ji's coefficients $a$ and $b$ owing to Eq.~\eqref{dictionary}. In the following, we will consider only the proton case and fix the renormalization scale to $\mu=2$ GeV.

\begin{figure}[t!]
	\centering
	\includegraphics[width=.95\textwidth]{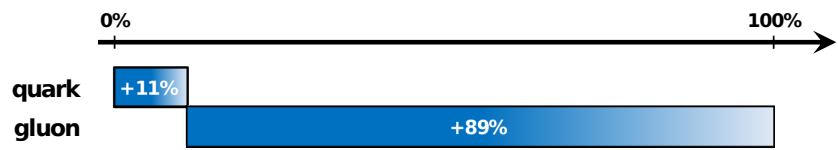}
\caption{\footnotesize{Trace decomposition of the nucleon mass~\eqref{tracedec} in units of $M$ at the scale $\mu=2$ GeV.}}
		\label{fig1}
\end{figure}

\begin{figure}[t!]
	\centering
	\includegraphics[width=.95\textwidth]{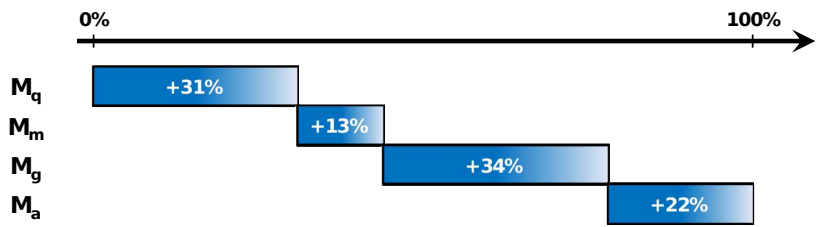}
\caption{\footnotesize{Ji's decomposition of the nucleon mass~\eqref{totalmass} in units of $M$ at the scale $\mu=2$ GeV.}}
		\label{fig2}
\end{figure}

\begin{figure}[t!]
	\centering
	\includegraphics[width=.95\textwidth]{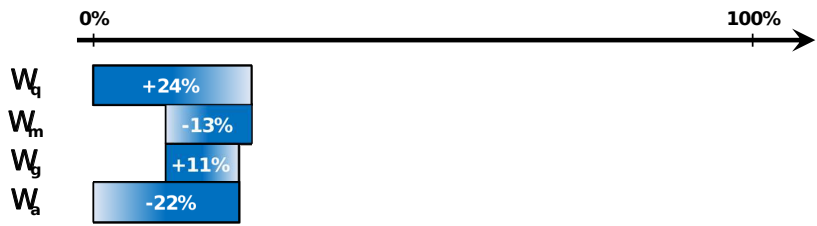}
\caption{\footnotesize{Decomposition of the nucleon pressure-volume work within Ji's picture~\eqref{Jipressure} in units of $M$ at the scale $\mu=2$ GeV.}}
		\label{fig3}
\end{figure}

\begin{figure}[t!]
	\centering
	\includegraphics[width=.95\textwidth]{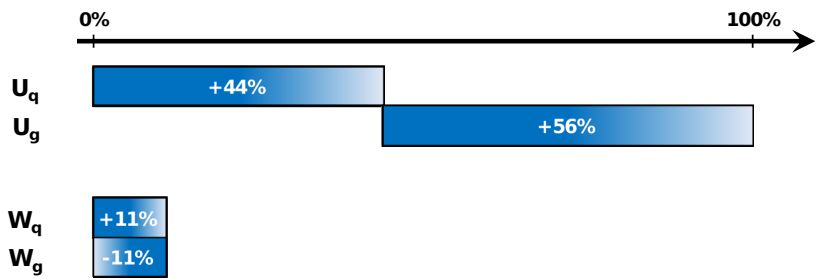}
\caption{\footnotesize{New decompositions of the nucleon mass and pressure-volume work~\eqref{newdecres} in units of $M$ at the scale $\mu=2$ GeV.}}
		\label{fig4}
\end{figure}

The parameter $a$, which is also interpreted as the average fraction of hadron momentum carried by quarks, can be extracted from deep-inelastic lepton-proton scatterings. A recent global analysis with leading-order parametrization obtained $a=0.546\pm 0.005$~\cite{Harland-Lang:2014zoa}. The parameter $b$ is related to the scalar charge of the proton and has been estimated to $b=0.113\pm 0.010$ by Gao {\it et al.}~\cite{Gao:2015aax}, based on a recent determination of the pion-nucleon $\sigma$-term~\cite{Hoferichter:2015dsa}, a recent lattice calculation of the strangeness content~\cite{Yang:2015uis} and neglecting the heavy quark contributions. It has also been suggested to extract the parameter $b$ from the trace anomaly using quarkonium-hadron scattering close to threshold~\cite{Kharzeev:1995ij,Kharzeev:1998bz}. For completeness, the anomalous quark mass dimension is approximatively given by $\gamma_m\approx -0.15$ for $n_f=3$ active flavors~\cite{Baikov:2014qja}. The various decompositions obtained with these values are depicted in Figs.~\ref{fig1}-\ref{fig4}.
\newline

In figure~\ref{fig1} we represent the trace decomposition given by Eq.~\eqref{tracedec} and determined only by the parameter $b$. As already discussed, it is largely dominated by the gluon contribution and is at the origin of the claim that most of the nucleon mass comes from gluons~\cite{Shifman:1978zn,Kharzeev:2002fm,Roberts:2016mhh,Roberts:2016vyn,Krein:2017usp}. The trace of the EMT being given by $T^\mu_{\phantom{\mu}\mu}=T^{00}-\sum_jT^{jj}$, what the quark and gluon contributions in Eq.~\eqref{tracedec} do actually represent are the combinations $U_{q,g}-3W_{q,g}$. Since the total pressure-volume work vanishes, one artificially supresses one of the contributions in favor of the other with pressure effects. As we will see below, it turns out that $W_q=-W_g>0$ which explains why the quark contribution appears to be much smaller than the gluon contribution. Because of these pressure effects, the sole trace of the QCD EMT does not provide enough information to determine the actual quark and gluon contributions to the hadron mass.

Ji's decomposition~\eqref{totalmass} is represented in Fig.~\ref{fig2}. The $M_q$ and $M_m$ contributions slightly differ from ones in the pie chart drawn in~\cite{Gao:2015aax} because we did not neglect the anomalous quark mass dimension $\gamma_m$. All the information about the hadron EMT in the forward limit, parametrized by Ji's coefficients $a$ and $b$, is included in this decomposition. As argued in the previous sections, a decomposition of the hadron mass into four contributions is however artificial. In the effective coupled two-fluid picture, the four contributions appear to be combinations of internal energies and pressure-volume works
\begin{align}
M_q&=\tfrac{3}{4}\,\tfrac{4+\gamma_m}{1+\gamma_m}\left(\tfrac{\gamma_m}{4+\gamma_m}\,U_q+W_q\right),\\
M_m&=\tfrac{1}{4}\,\tfrac{4+\gamma_m}{1+\gamma_m}\left(U_q-3W_q\right),\\
M_g&=\tfrac{3}{4}\left(U_g+W_g\right),\\
M_a&=\tfrac{1}{4}\left(U_g-3W_g\right).
\end{align}
In order to avoid adding apples and oranges, Ji required that each term must represent some internal energy, which amounts to adopting an effective coupled four-fluid picture. In such a picture, no new information arises from the corresponding decomposition of the pressure-volume work~\eqref{Jipressure} represented in Fig.~\ref{fig3}, since the individual contributions are directly determined by the energy ones
\begin{equation}\label{JiEOS}
\mathcal W_q=\tfrac{1}{3}\left(M_q+\tfrac{12}{4+\gamma_m}\,M_m\right),\qquad
\mathcal W_m=-M_m,\qquad
\mathcal W_g=\tfrac{1}{3}\,M_g,\qquad
\mathcal W_a=-M_a,
\end{equation}
as a consequence of the conventional splitting into traceless and trace parts\footnote{Note that the appearance of $M_m$ in the expression for $\mathcal W_q$ comes from the reshuffling the quark mass terms between the quark traceless and trace parts~\eqref{reshuffling}.}. In the effective coupled two-fluid picture, they are explicitly given by
\begin{align}
\mathcal W_q&=\tfrac{1}{4}\,\tfrac{4+\gamma_m}{1+\gamma_m}\left(U_q-\tfrac{8-\gamma_m}{4+\gamma_m}W_q\right),\\
\mathcal W_m&=-\tfrac{1}{4}\,\tfrac{4+\gamma_m}{1+\gamma_m}\left(U_q-3W_q\right),\\
\mathcal W_g&=\tfrac{1}{4}\left(U_g+W_g\right),\\
\mathcal W_a&=-\tfrac{1}{4}\left(U_g-3W_g\right).
\end{align}
The information about the energy content and the stability of the system are a priori independent. By treating the traceless and trace parts as separate EMT, Ji effectively combined both information into a single decomposition. While mathematically correct, such a decomposition is however not satisfactory from the physical point of view because gluons involved in the trace anomaly are treated as if they were distinguishable from those carrying kinetic and potential energies (they are indeed associated with different equations of state in Ji's approach).


Since hadrons are made of quarks and gluons, it seems more natural to decompose into two contributions only. Ji's coefficients $a$ and $b$ appear to characterize two different hadron properties of the hadron, namely mass and pressure-volume work. Accordingly, instead of a single decomposition of the hadron mass into four contributions like Ji's, we propose to consider separately the decompositions of hadron mass and pressure-volume work into two contributions, given by Eq.~\eqref{newdecres} and represented in Fig.~\ref{fig4}. We do not require any splitting into traceless and trace parts, and hence we do not fix a priori the relation between energy density and pressure. Note that all the standard energy conditions are satisfied~\cite{Visser:1999de,Curiel:2014zba}
\begin{equation}
\begin{aligned}
\text{null energy condition}&\qquad&\epsilon_i+p_i&\geq 0,\\
\text{weak energy condition}&&\epsilon_i&\geq 0\quad\text{and}\quad\epsilon_i+p_i\geq 0,\\
\text{dominant energy condition}&&\epsilon_i&\geq|p_i|,\\
\text{strong energy condition}&&\epsilon_i+p_i&\geq 0\quad\text{and}\quad\epsilon_i+3p_i\geq 0.
\end{aligned}
\end{equation}
One is of course free to define quark mass and trace anomaly contributions, but these turn out to be purely conventional and to lead to physical conumdrums when treated covariantly.

\section{Conclusions}\label{sec5}

We used forward matrix elements of the energy-momentum tensor to characterize static mechanical properties of hadrons. The components of such an energy-momentum tensor can be interpreted semi-classically in terms of parton energy density and pressure averaged over time and the hadron proper volume. 

We showed that, because of pressure effects, the physical interpretation of the standard decompositions of the hadron mass have to be considered with care. In the trace decomposition, a pressure-volume work contribution artificially emphasizes the role played by gluons. In Ji's decomposition, the splitting of the quark and gluon energy-momentum tensors into traceless and trace parts mixes the information about the hadron mass budget with the pressure-volume work budget. This can be understood by the fact that Lorentz symmetry is already used to provide a physical interpretation of the various components of the energy-momentum tensor. It cannot be used a second time to define separate mass contributions without modifying the physical picture. In particular, from the point of view of physical interpretation, using Ji's decomposition amounts to treating gluons involved in the trace anomaly and those carrying kinetic and potential energies as separate entities with different equations of state, which is not physically acceptable.

Since hadrons are made of quarks and gluons, it is natural to decompose their mass into two contributions only. Any further decomposition that is not based on the properties of the constituents will be somewhat arbitrary and will mix internal energy with pressure-volume work. This mixing is mathematically harmless because the total pressure-volume work vanishes for a stable system, but it is a problem for the physical interpretation of the individual contributions.

We proposed a new picture, where the hadron mass and pressure-volume work budgets are kept separate and expressed in terms of the sole quark and gluon contributions which are physically unambiguous. In particular, unlike Ji's decomposition we do not fix a priori the equations of state for quarks and gluons. 

Finally, we quantitatively compared the different decompositions based on recent phenomelogical estimates. It turned out that, as expected for highly relativistic systems, pressure-volume work contributions are of the same order of magnitude as internal energy contributions. In particular, quarks are responsible in average for the repulsive force and gluons for the attractive force inside nucleons. 
\newline

Having clearly identified the pressure contributions opens many interesting applications related to compact stars. For example, determining the quark and gluon equations of state inside a nucleon may give important clues about the internal structure of compact stars.

\section*{Acknowledgement}

This work is a result of discussions held at the workshop ``The Proton Mass: At the Heart of Most Visible Matter'' at the ECT* Trento, on 3-7 April 2017. This work has been supported by the Agence Nationale de la Recherche under the project ANR-16-CE31-0019.

\end{document}